\documentclass[a4paper,10pt,twoside]{cpc-hepnp}
\usepackage{CJK,upgreek,fancyhdr}
\usepackage{multicol}
\usepackage{graphicx}
\usepackage{booktabs}
\usepackage{graphics}
\usepackage{amssymb,bm,mathrsfs,bbm,amscd}
\usepackage[tbtags]{amsmath}
\usepackage{lastpage}
\usepackage{color}
\usepackage{bm}
\usepackage{ulem}

\begin{document}
\begin{CJK*}{GB}{gbsn}

\fancyhead[c]{\small Submitted to Chinese Physics C}
\fancyfoot[C]{\small 010201-\thepage}

%\footnotetext[0]{Received 31 June 2015}

\title{A possible explanation of the knee of cosmic light component spectrum from 100 TeV to 3 PeV\thanks{ supported by the National Natural Science Foundation
of China (NSFC 11433004, 11363006, 11103016, 11173020), Top Talents Program of Yunnan Province(2015HA030) and the Natural Science Foundation of Yunnan Province(2015FB103). }}

\author{%
\quad Wen-Hui Lin(林文慧)$^{1}$%
\quad Bi-Wen Bao(鲍必文)$^{1}$
\quad Ze-Jun Jiang(姜泽军)$^{1}$ \email{zjjiang@ynu.edu.cn}%
\quad Li Zhang (张力)$^{1}$ \email{lizhang@ynu.edu.cn}%
}

\maketitle

\address{%
$^1$ Department of Astronomy, Yunnan University, Key Laboratory of Astroparticle physics of Yunnan Province,
      Kunming 650091, China.\\

}
%$^2$ {\bf Example}: Institute of High Energy Physics, Chinese Academy of Sciences, Beijing 100049, China\\}

\begin{abstract}
The mixed Hydrogen and Helium (H + He) spectrum with a clear {steepening} at $\sim 700$ TeV has been detected by ARGO-YBJ experiments. In this paper,  {we demonstrate that} the observed H + He spectrum {can be well} reproduced {with} the model {of} cosmic rays escaping from the supernova remnants (SNRs) in our Galaxy. In this model, particles are accelerated in a SNR through a non-linear diffusive shock acceleration mechanism and three components of high energy light nuclei escaped from the SNR are considered. {It should be noted that the proton spectrum observed by KASCADE can be also explained by this model given a higher acceleration efficiency.}
\end{abstract}


\begin{pacs}
%1---3 PACS codes (Physics and Astronomy Classification Scheme, http://www.aip.org/pacs/pacs.html/)
98.38.Mz, 98.58.Db, 95.85.Pw
\end{pacs}

\footnotetext[0]{\hspace*{-3mm}\raisebox{0.3ex}{$\scriptstyle\copyright$}2013
Chinese Physical Society and the Institute of High Energy Physics
of the Chinese Academy of Sciences and the Institute
of Modern Physics of the Chinese Academy of Sciences and IOP Publishing Ltd}%

\begin{multicols}{2}

\section{Introduction}

%This template is based on article.cls with some modifications. The character size is set to 10 pt and the paper size is A4. The corresponding class file is cpc-hepnp.cls~\cite{lab1}.

%This template uses the CJK package to display Chinese, Japanese and Korean author names. If you do not wish to display characters from any of these languages, and find that LaTeX compilation fails because you are missing this package, please delete all references to the CJK package from this template. (It is referenced in the \texttt{usepackage\{CJK\}} line and at the beginning and end of the document.)

%Some ten macro packages are used in this template\cite{lab1,lab2,lab3}. All of these macro packages are built-in Medium tex, and do not need to be downloaded or installed. Latex 2.3 or higher versions can be used. You can add more macro packages if needed while writing, but attention should be paid to the possible conflicts of macro packages.

%You can freely use a variety of tex instructions or redefine some of their instructions to achieve your desired results. However, we still have some special regulations which are mentioned in the following.

{Supernova remnants (SNRs) are generally believed to be the origin of Galactic cosmic rays (CRs) \cite{lab26,lab27, lab28, lab29, lab30}.} The knee of $\sim 3\times 10^{15}$ eV as a feature coinciding with the maximum energy of the light component of cosmic rays and {the} transition to a gradually heavier mass composition {are} mainly based on KASCADE results \cite{lab1}. Some recent data, however, appear to {challence} this finding: the combined detection of showers with a wide field of view Cherenkov telescope (WFCT) and ARGO-YBJ find a flux reduction in the light component at $\sim$ 700 TeV \cite{lab2}, {a factor ${\sim 0.2}$ of the former.} This observed result favors SNR's origin in our Galaxy and help us understand the acceleration mechanism inside the SNRs.

Diffusive shock acceleration (DSA) is assumed to take place inside a SNR as an efficient particle acceleration mechanism {\cite{lab31,lab32, lab33, lab34}}. However, a high acceleration efficiency implies strong coupling between the accelerated particle population, the shock structure, and the electromagnetic fluctuations, resulting in a non-linear diffusive shock acceleration (NLDSA) \cite{lab1,lab3, lab4}. To solve NLDSA problem, three approaches have been proposed: kinetic semi-analytic solutions, Monte Carlo numerical simulations of the full particle population, and fully numerical simulations. The results obtained with the three different techniques are consistent with each other in terms of both accelerated spectra and hydrodynamics (see the review of ref. \cite{lab6}). We will base on the kinetic semi-analytical model given by Ref. \cite{lab8} for its computational convenience.

In this paper, we will investigate the properties of particle spectra efficiently accelerated in Galactic SNRs and try to {provide} a possible explanation of the  position of knee of light component cosmic {ray} spectrum observed by ARGO-YBJ \cite{lab2} and KASCADE \cite{lab21}, \cite{lab22}, \cite{lab23}. The paper is structured as follows. We give a brief review of the NLDSA model of Ref \cite{lab8} in section 2, and apply this model to explain the observed light component cosmic {ray} spectra in section 3. Finally, a summary is given in Section 4.

\section{The Review of NLDSA Model}

In the NLDSA model given by Ref. \cite{lab8}, {the shock is placed at a distance ${x=0}$, so the upstream and downstream region correspond to ${x<0}$ and ${x>0}$, respectively.} Physical quantities measured at upstream infinity, {immediate} upstream of the shock, and downstream are labelled by subscripts 0, 1, and 2, respectively. Generally, it is convenient to define two different compression ratios for subshock and total shock \cite{lab5}: ${r}_{\rm sub}= {\tilde{u}_1}/\tilde{u}_2$ and ${r}_{\rm tot}= \tilde{u}_0/\tilde{u}_2$, where $\tilde{u}=u+ u_A$, $u$ is the bulk plasma velocity in the shock frame, {and} $u_A$ is the Alfv\'{e}n velocity with respect to the background plasma.

In such a system, a diffusive-convection equation describing the transport of the $i$th particles in the shock frame can be expressed as
\cite{lab8}:
\begin{eqnarray}
\label{eq:difftran}
\tilde{u}\frac{\partial f_i}{\partial x} =
\frac{p}{3}\frac{\partial \tilde{u}}{\partial x}\frac{\partial f_i}{\partial p}+
\frac{\partial }{\partial x}\left[\kappa_i(x,p)\frac{\partial f_i}{\partial x} \right] + Q_i(x,p)\;,
\end{eqnarray}
where  subscript $i$ represents the $i$th particles, $f_i=f_i(x, p)$ is the distribution {function} of the $i$th particles, $\kappa_i(x,p)$ is the Bohm-like parallel diffusion coefficient with a magnetic field strength $B(x)$ and is given by
\begin{equation}
\kappa_i(x,p)=\frac{u(x)}{3}r_{\rm L}(x,p) = \frac{u(x)pc}{3Z_i eB(x)}\;,\label{kappa}
\end{equation}
$Q_i(x,p)$ is the $i$th particle injection term and is given by
\begin{eqnarray}\label{Qxp}
  Q_i(x,p) = \eta_i \frac{n_{0}u_{0}}{4\pi p_{{\rm inj}, i}^{2}}\delta(p-p_{{\rm inj}, i})\delta(x)\;.
\end{eqnarray}
where $\eta_i$ is the fraction of the $i$th particles crossing the shock injected in the acceleration process, $p_{{\rm inj}, i}=Z_{i}~p_{\rm inj, \rm H}$ is the injection momentum of the $i$th particles, and $\delta(x)$ is {the} position {where} particles {are injected} at the shock.
In the thermal-leakage model {\cite{lab8,lab10}},
 \begin{eqnarray}
&p_{{\rm inj}, \rm H}= \xi_{\rm H}\sqrt{2m_{\rm H} k_{\rm B}T_{\rm H, 2}}\;,\\
&T_{\rm H, 2} = T_0(R_{\rm tot}/R_{\rm sub})^{\gamma-1}\frac{\gamma+1-(\gamma-1)R_{\rm sub}^{-1}}{\gamma+1-(\gamma-1)R_{\rm sub}}\;,
\end{eqnarray}
where $k_{\rm B}$ is the Boltzmann constant, $m_{\rm H}$ is proton mass, {${\gamma=5/3}$ is the ratio of specific heats of gas} and $\xi_{\rm H}$ is a parameter which defines $p_{\rm inj, \rm H}$ as a part of the momentum of the thermal protons in the downstream region.

In the absence of the dynamical back-reaction of the accelerated particles and magnetic field amplification at the shock, Eq.~(\ref{eq:difftran}) can {be} independently solved. In other words, the pressure of the accelerated particles $P_{\rm CR}(x)$ and magnetic field pressure $P_{\rm B}(x)$ are negligible in comparison with the pressure $P_{\rm g}(x)$ of the gas with a density $\rho$. In this case (called test particle approximation), the conservation equations of mass, momentum and energy across the shock have the trivial solutions: $\rho =$ constant, $u =$ constant, and $P_{\rm g} =$ constant.

In a general case, {the} two non-linear effects {above} must be considered.

\textit{(i) the dynamical back-reaction of the accelerated particles.} Because the particles are efficiently accelerated at shock, the pressure $P_{\rm CR}$ of the accelerated particles must be included in the energy conservation equation, which results in the slow down of upstream plasma velocity in the shock frame, forming a  so-called dynamical shock precursor. The {normalized} pressure of the accelerated particles can be estimated as
\begin{eqnarray}
\label{eq:PCR}
 P_{{\rm CR}}(x) = \frac{4\pi}{3\rho_{0}u_{0}^{2}} \sum_i \int_{p_{{\rm inj}, i}}^{\infty}dp~ p^{3}\tilde{u}(x)f_i(x,p)\;.
\end{eqnarray}

\textit{(ii) the effect of the magnetic field amplification.} Here magnetic field amplification due to the streaming instability of plasma flow is considered. In this case, the streaming instability of plasma flow will lead to the magnetic field amplification, and then the magnetic pressure ($P_{\rm B}$) will have a significant change to affect shock compression ratios and particle's spectra. Following Ref. {\cite{lab8}}, the {normalized} pressure of the magnetic field can be expressed as
\begin{eqnarray}\label{eq:PB}
P_{\rm B}(x)=\frac{2}{25}\frac{[1-U(x)^{5/4}]^2}{U(x)^{3/2}}\;,
\end{eqnarray}
where $U(x)=(\rho/\rho_0)[u^2(x)/u^2_0]$. And then the amplified magnetic field is estimated as
\begin{equation}
B(x)=\sqrt{8\pi P_B(x)}\;.
\end{equation}

Considering both of the effects mentioned above, {the momentum conservation equation, normalized to ${\rho_{0}u_{0}^{2}}$, can be represented as}:
\begin{eqnarray}
U(x)+P_{\rm CR}(x)+P_{\rm B}(x)+P_{\rm g}(x)=1+\frac{1}{\gamma M_0^2}\;.\label{eq:ux}
\end{eqnarray}
where $P_{\rm CR}(x)$ and $P_{\rm B}(x)$ are given by Eqs.~(\ref{eq:PCR}) and~(\ref{eq:PB}), respectively, and $P_{\rm g}(x)=\frac{U(x)^{-\gamma}}{\gamma M_0^2}$ is the normalized pressure of the background gas with adiabatic
index $\gamma$, {${M_{0}=\frac{u_{0}}{c_{s}}}$ is the Mach number of the fluid at upstream infinity.}

Equation (\ref{eq:difftran}) with the spatial boundary condition $f_i(x_0, p) = 0$ has been solved and the solution can be expressed as \cite{lab5}
\begin{eqnarray}\label{fxp}
 f_{\rm i}(x,p)=f_{\rm sh, i}(p)e^{-\int_{x}^{0} dx'\frac{\tilde{u}(x')}{\kappa_i(x',p)}\left[ 1-\frac{W_{i}(x,p)}{W_{i,0}(p)}\right]}\;.
\end{eqnarray}
where $f_{\rm sh,\rm i}(p)$ is the distribution function at shock, which is
\begin{eqnarray}\label{eq:hnsolshock}
f_{\rm sh,\rm i}(p)=\frac{\eta_{\rm i} n_{0} q_{\rm p,\rm i}(p)}{4\pi p_{\rm inj,\rm i}^{3}}
	e^{\left\{-\int_{p_{\rm inj,\rm i}}^{p}\frac{dp'}{p'}
	q_{\rm p,\rm i}(p')\left[ U_{\rm p, i}(p')+\frac{1}{W_{i, 0}(p')}\right]\right\}}\;.
\end{eqnarray}
where $U_{\rm p, i}(p)=U_1-\int_{x_0}^{0} dx [dU(x)/dx][f_i(x,p)/f_{\rm sh, i}(p)]$ and
$q_{\rm p, i}(p)=3R_{\rm tot}/(R_{\rm tot} U_{\rm p, i}(p)-1)$; the function $W_{i}(x,p)$ in Eq. (\ref{fxp}) is given by
\begin{eqnarray}\label{Wxp}
W_{i}(x,p)=\int_{x}^{0} dx'\frac{u_0} {\kappa_i(x',p)}e^{\int_{x'}^{0}dx''\frac{\tilde{u}(x'')}{\kappa_i(x'',p)}}\;.
\end{eqnarray}
and $W_{i,0}(p)=W_{i}(p)|_{x=x_0}$. The escape flux can be written by
\begin{eqnarray}\label{Phixp}
\Phi_{\rm esc,\rm i}(p)=-\kappa_i(x,p)\left. \frac{\partial f_{\rm i}}{\partial x}\right|_{x_0}=- \frac{u_0f_{\rm sh,\rm i}(p)}{W_{\rm i,0}(p)}\;.
\end{eqnarray}

Therefore, Eq.~(\ref{eq:difftran}) is coupled with the equations describing mass, momentum, and energy flux conservations, leading to the problem of NLDSA whose solution can be obtained through the iterative method described in Refs. \cite{lab5} and \cite{lab12}.

\section{Particle Injection from a SNR}

To perform our calculation, {the} following assumptions about a SNR's evolution are made: the SNR is produced in a supernova explosion with an energy  $E_{\rm SN}=10^{51}$ erg and an ejecta mass $M_{\rm ej}=1.4M_\odot$,  and a shock moves with velocity $u_0 = 4000$ km/s in a homogeneous and hot medium with a numer density $n_0=0.01$ cm$^{-3}$, a temperature $T_0 = 10^6$ K and a background magnetic field $B_0 = 5~\mu$G. According to the analytical recipe given in Ref. \cite{lab9}, the SNR evolution is divided into two stages{\cite{lab8}}:
\begin{itemize}
\item[(1)] Ejecta-dominated stage with $\tau= t/T_{\rm ST} \leq 1$, both radius and velocity of the SNR are given by
\begin{eqnarray}
&R_{\rm sh}(t) \simeq 14.1 \tau^{4/7} \;\;{\rm pc}\;,\\
&V_{\rm sh}(t) \simeq 4140 \tau^{-3/7}\;\; {\rm km/s}\;,
\end{eqnarray}
where $T_{\rm ST} \simeq 2000$ yr is used;
\item[(2)] Sedov - Taylor stage with $\tau = t/T_{\rm ST} \geq 1$,
\begin{equation}
R_{\rm sh}(t) \simeq 16.2 (\tau-0.3)^{2/5}\;\;{\rm pc}\;,
\end{equation}
\begin{equation}
V_{\rm sh}(t) \simeq 3330 (\tau-0.3)^{-3/5}\;\;{\rm km/s}\;.
\end{equation}
\end{itemize}

Adiabatic loss is included because of the shell expansion. The energy $E(t)$ of a particle with energy $E_{0}$ advected downstream at time $t_{0}$ is given by \cite{lab8}
where $4/3\leq \gamma \leq 5/3$.

The accelerated particles can escape from a shell volume $V=4\pi R^2_{\rm sh}d R_{\rm sh}$, where $dR_{\rm sh}=V_2(t) dt=[V_{\rm sh}(t)/\bm{r}_{\rm tot}]dt$ and $dt$ is the time {increment}. Since the particle numbers per unit volume per unit energy can be expressed as $J_{\rm i}(E, t)=4\pi p^2 f_{\rm sh,i}(p)dp/dE$, where $f_{\rm sh,i}(p)$ is the distribution function at shock radius $R_{\rm sh}(t)$, the particle numbers per unit energy in the shell volume $V$ can be estimated as $J_{\rm i}(E, t)\times V$. Because the shell volume evolves with time $t$ during SNR evolution, the particle numbers per unit energy {are}

\begin{equation}
\phi_{\rm i}(E)=4\pi \int^{T_f}_{T_i}J_{\rm i}(E, t)R^2_{\rm sh}(t)[V_{\rm sh}(t)/\bm{r}_{\rm tot}]dt\;,
\end{equation}

where $T_i$ and $T_f$ are the initial and final times of the SNR evolution, here $T_i= 0.1T_{\rm ST}$ and $T_f=15T_{\rm ST}$ with $T_{\rm ST}=2000$ yr are used. There are three kinds of components for the particles escaping from the SNR \cite{lab5}:

(i) The numbers per unit energy of particles which instantaneously escape around a maximum momentum $p_{\rm max}(t)$ from the upstream free escape boundary at $x=x_0$ during the Sedov - Taylor (ST) stage, where $p_{\rm max}(t)$ is determined by the finite size of the
SNR during the ST stage, which can be estimated by $\kappa_i(p_{\rm max})/V_{\rm sh}(t)\approx \chi R_{\rm sh}(t)$, {assuming} $x_0$ to be a fraction $\chi$ of the radius $R_{\rm sh}(t)$ of the SNR with a shock velocity $V_{\rm sh}(t)$. In this case, the number density of particles per unit energy is $4\pi p^2(\Phi_{\rm esc,\rm i}(p)/V_{\rm sh}(t_0))dp/dE$, where $\Phi_{\rm esc,\rm i}(p)$ is given by Eq. (\ref{Phixp}), therefore the particle numbers per unit energy are
\begin{eqnarray}\label{eq:phiesc1}
q_{\rm esc, i}(E_0)= \frac{16\pi^2}{c^2}\int^{T_f}_{T_i}pR_{sh}^{2}(t)E_0\frac{\Phi_{\rm esc,\rm i}(p)}{\bm{r}_{\rm tot}}dt\;.
\end{eqnarray}

(ii) The numbers per unit energy of particles which are advected in the downstream region, leading to adiabatic losses
as a consequence of the shell expansion, where the particles can escape at $p > p_{\rm esc}(t)$  at any given time and $p_{\rm esc}(t)$ can be estimated by $\kappa_i(p_{\rm esc}, B_2)/V_2=x_0$ with $V_2=V_{\rm sh}/\bm{r}_{\rm tot}$. In this case, the particle numbers per unit energy {are}
\begin{eqnarray}\label{eq:phiesc2}\nonumber
q_{\rm adv, i}(E_0)&=& \frac{16\pi^2}{c^2}\int^{T_f}_{T_i}pR_{sh}^{2}(t)E_0f_{\rm sh,i}(p)\frac{V_{\rm sh}(t_0)}{\bm{r}_{\rm tot}}\\
&&\times \left(\frac{V_{\rm sh}(t)}{V_{\rm sh}(t_0)}\right)^{2/3\gamma}dt\;.
\end{eqnarray}

(iii) The numbers per unit energy of particles which escape the acceleration region from a broken shell at the end of a SNR's evolution,
\begin{eqnarray}\label{eq:phiesc3}
q_{\rm shell, i}(E_0)= \lambda \times\frac{16\pi^2}{c^2}\int^{T_f}_{T_i}pR_{sh}^{2}(t)E_0f_{\rm sh,i}(p)\frac{V_{\rm sh}(t_0)}{\bm{r}_{\rm tot}} dt\;,
\end{eqnarray}
where the fraction of downstream escaping particles is taken as $\lambda\approx $ 10\%.

Therefore, the numbers per unit energy  of the $i$th particles escaping from a single SNR is the sum of Eqs. (\ref{eq:phiesc1}) - (\ref{eq:phiesc3}), i. e.,
\begin{eqnarray}\label{eq:phii}
q_i(E)=  q_{\rm esc, i}(E) + q_{\rm adv, i}(E)+ q_{\rm shell, i}(E)\;.
\end{eqnarray}

As an example, the {spectra of} H and He {nuclei} injected into the interstellar space are shown in Fig.~\ref{fig:1}. The model parameters are as follows: ${\xi_{H} = 3.0}$, $\chi=0.5$ $T_0 = 10^6$ K, $n_0 = 0.01$ cm$^3$, and $B_0=5~\mu$G. In this figure, the spectra of three components mentioned above for H and He nuclei have been shown. From the figure, the first component dominates high-energy end, and the second component is a main contributor at lower energy. {Moreover, from the bell-shaped curves (dotted line) which the particles have escaped to SNR from the upstream boundary, we can see that escaping occurs at highest energies.} On the other hand, the cut-off energy of He is larger than that of H. {For parameters used here, the shock is modified by the accelerated particles: ${r_{\rm sub}=3.41}$ and ${r_{\rm tot}=4.73}$.}

\begin{center}
\includegraphics[width=9cm]{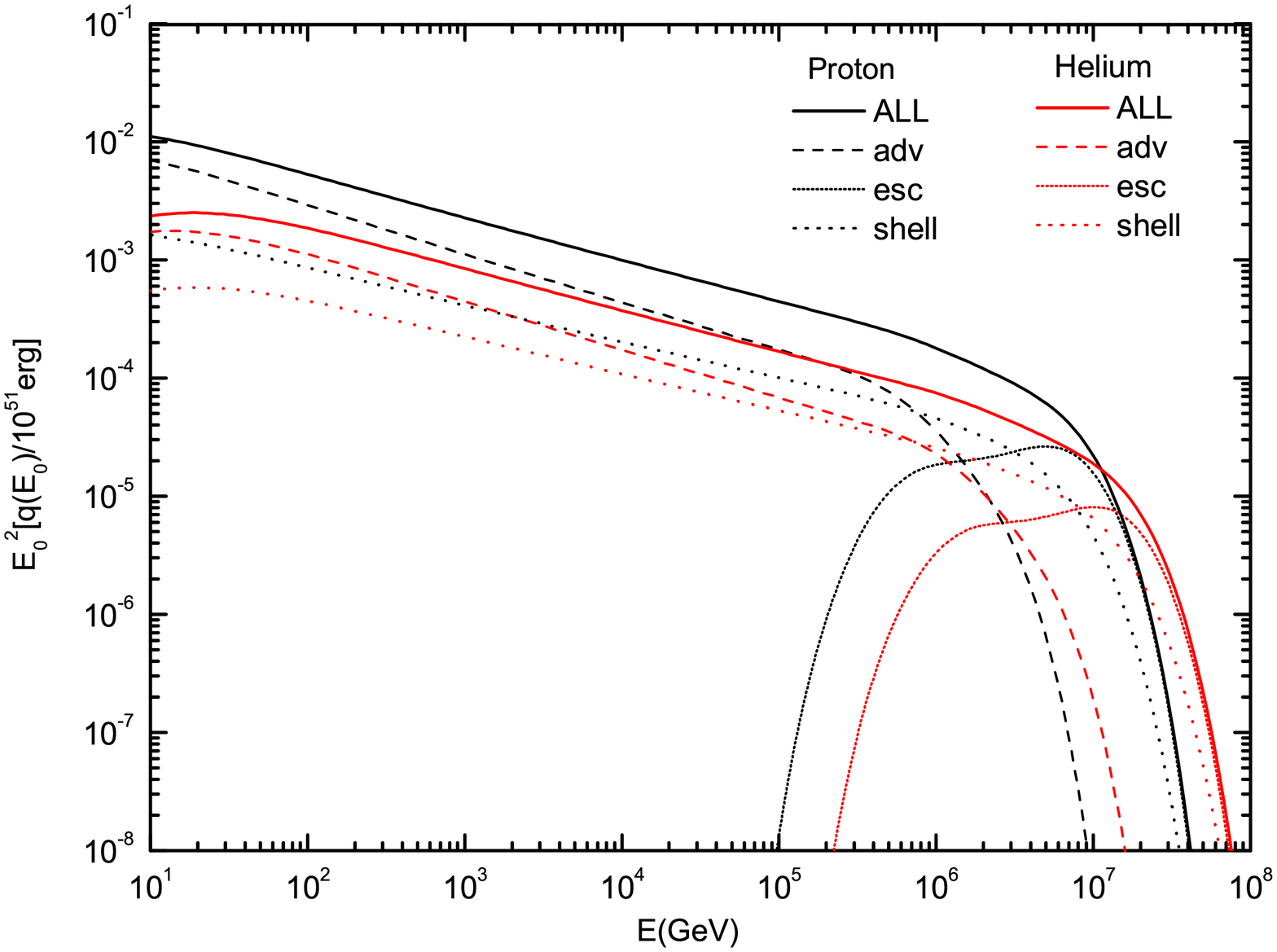}
\figcaption{\label{fig:1}  Escape spectra of protons and He nuclei from a single SNR. Three components of the escape spectra for protons (black line) and He (red line) nuclei are presented by long dashed, dotted, and short dashed lines, the black  and red solid lines represent total spectra of protons and He nuclei, respectively. The model parameters are  $\xi_{\rm H} = 3.0$, $\chi=0.5$, $T_0 = 10^6$ K, $n_0 = 0.01$ cm$^{-3}$, and $B_0=5~\mu$G. }
\end{center}

Note that one parameter has an important influence on the CR spectrum, i.e., the parameter $\xi_{\rm H}$ which describes the acceleration efficiency.  {In Fig. \ref{fig:2}, it can be seen that the lower acceleration efficiency (i.e., a larger ${\xi_{\rm H}}$) the particles, the flatter the resulting spectra and smaller the maximum energy.}

\section{Proton and He spectra Observed at the Earth}

In this section, the spectra of H and He nuclei observed at the Earth are calculated. It is assumed that the propagation can be approximated by a simple leaky box model with 3 percent of SN explosion rate in our Galaxy. In this case, the energy spectrum $N_i(E)$ of the $i$th particles observed at the Earth is given by \cite{lab13}
\begin{eqnarray}\label{}
  N_i(E) \propto q_i(E)\left(\frac{1}{\lambda_{\rm esc, i}}+\frac{1}{\lambda_{\rm int,i}}\right)^{-1}\;,
\end{eqnarray}
where the escape path length $\lambda_{\rm esc, i}$ is a function of the particle magnetic rigidity $R_i = pc/Z_i$ and is approximated as
$\lambda_{\rm esc} = 7.3(R_i/10~{\rm GV})^{-\delta}\beta(p)$ g/cm$^2$, where $\delta =0.3-0.6$, $\beta(p)$ is the dimensionless speed of a nucleus of momentum $p$, $\lambda_{\rm int,i}$ is the interaction length and $\lambda_{\rm int,i}= \lambda_{0,\rm i}(E/10~\rm GeV)^{-\epsilon_i}$, where $\lambda_{0, \rm H}=\lambda_{0, \rm He}=50$ g cm$^{-2}$, $\epsilon_{\rm H}=0.05$, $\epsilon_{\rm He}=0.0416$ \cite{lab13}.

\begin{center}
\includegraphics[width=9cm]{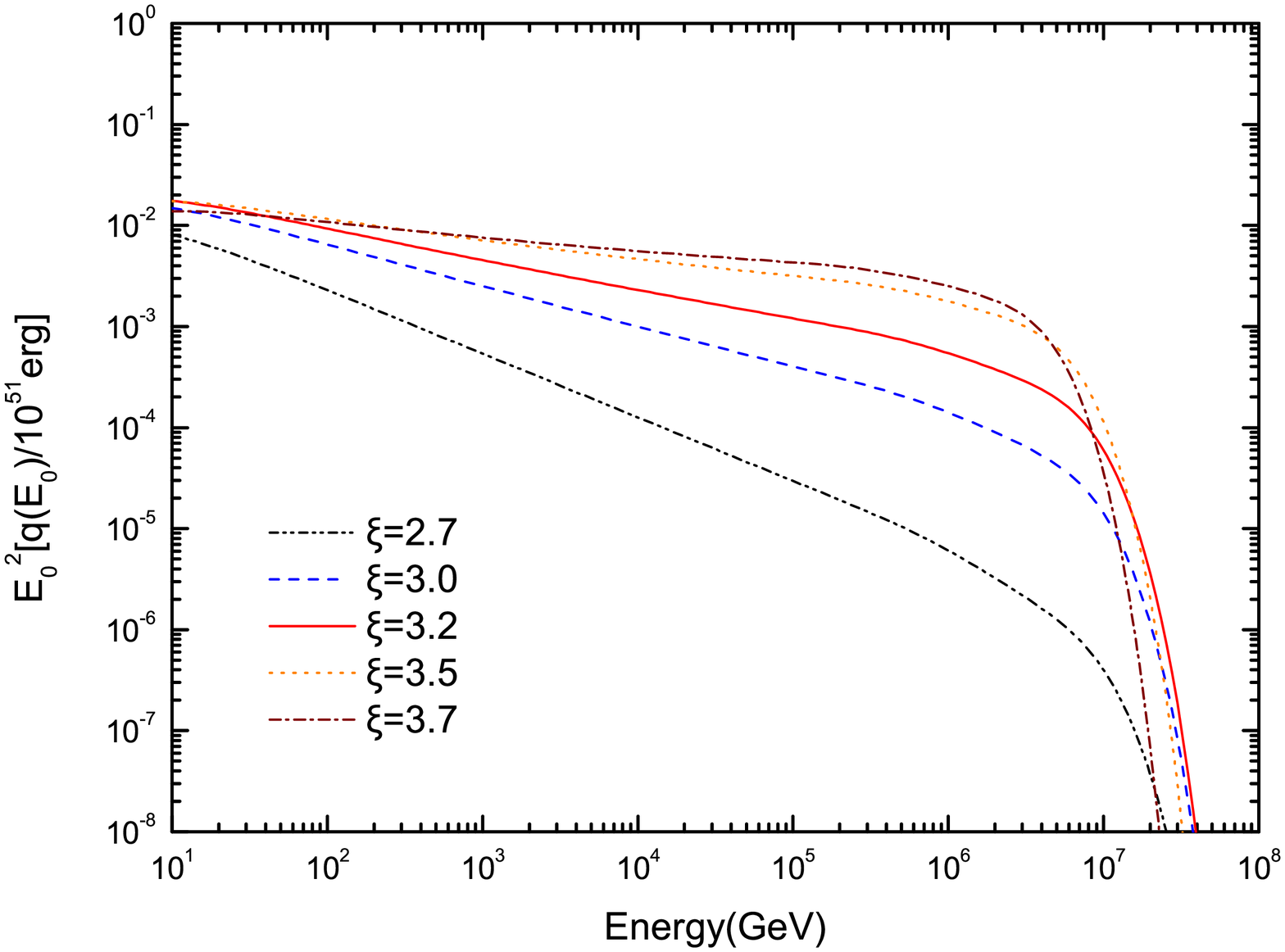}
\figcaption{\label{fig:2} Cumulative proton spectra for $\xi_{H}=2.7$ (dash-dot-dot line), $\xi_{H}=3.0$ (dash line), $\xi_{H}=3.2$ (solid line), $\xi_{H}=3.5$ (dot line) and $\xi_{H}=3.7$ (dash-dot line).}
\end{center}

In order to estimate the value of $\eta_i$ from the measurement, following Ref. \cite{lab7}, the ratio of abundances between ions and protons at the same momentum $p*= 10^5$ GeV/c measured at the Earth is defined as $K_{i\rm H}=n_i/n_{\rm H}$, and $\eta_i/\eta_{\rm H}\approx K_{i\rm H}Z^{-(\delta +\beta -3)}_i$. {The} distribution function is assumed to be a power law with a slope $\beta$, which can be obtained in the test particle approximation. {Here} $\beta +\delta=4.7$ is used \cite{lab7}, so $\eta_{\rm He}/\eta_{\rm H}\approx 0.31 K_{\rm He~H}$.

Figure \ref{fig:3} shows the comparison of our model results with the H + He flux observed by ARGO-YBJ experiments \cite{lab2,lab14,lab15}. The H + He spectrum has a knee feature of $\sim 700$ TeV \cite{lab2}. For comparison, the observed data from various experiment groups are also shown. It can be seen from the figure that our model results can reproduce the flux observed by ARGO-YBJ experiments well. Moreover, the knee feature can be explained by the sum of H and He spectra but He spectrum dominates the high energy end. {Note that the acceleration is efficient in this case, ${\xi_{\rm H}=3.8}$ corresponds to ${\eta_{\rm H}\approx 6.5\times 10^{-5}}$ and ${\eta_{\rm He}\approx 0.31 K_{\rm He~H}\times\eta_{\rm H}\sim 2.01\times 10^{-5}}$.} And the deviation at low energy may be due to solar modulation, {and} we did not consider the factor in our model.

\begin{center}
\includegraphics[width=9cm]{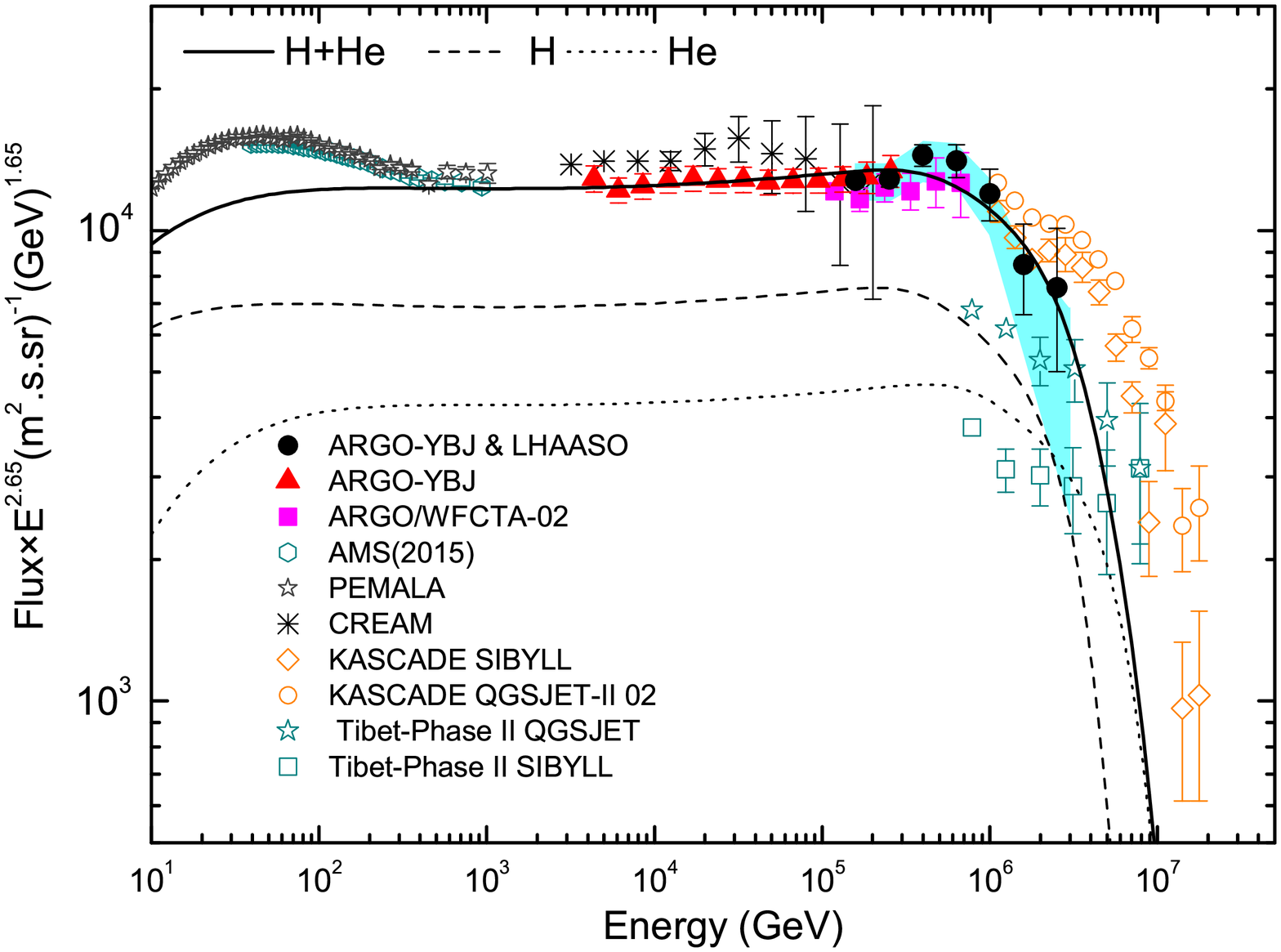}
\figcaption{\label{fig:3}  Comparison of our model results with the observed H + He flux. The observed data are taken from ARGO-YBJ\cite{lab2,lab14,lab15}, AMS \cite{lab16, lab17}, PAMELA \cite{lab18}, CREAM \cite{lab19}, the hybrid experiment\cite{lab15} below the knee, Tibet $\rm AS_{\gamma}$\cite{lab20}, and KASCADE \cite{lab21,lab22,lab23} above the knee. Long dashed and short dashed lines represent predicted fluxes of H and He components, respectively, and the solid line is the flux sum of two components. Model parameters are $\xi_{\rm H} = 3.8$, $\delta =0.55$, $\chi=0.5$, $T_0 = 10^6$ K, $n_0 = 0.01$ cm$^{-3}$, and $B_0=5~\mu$G. }
\end{center}

\begin{center}
\includegraphics[width=9cm]{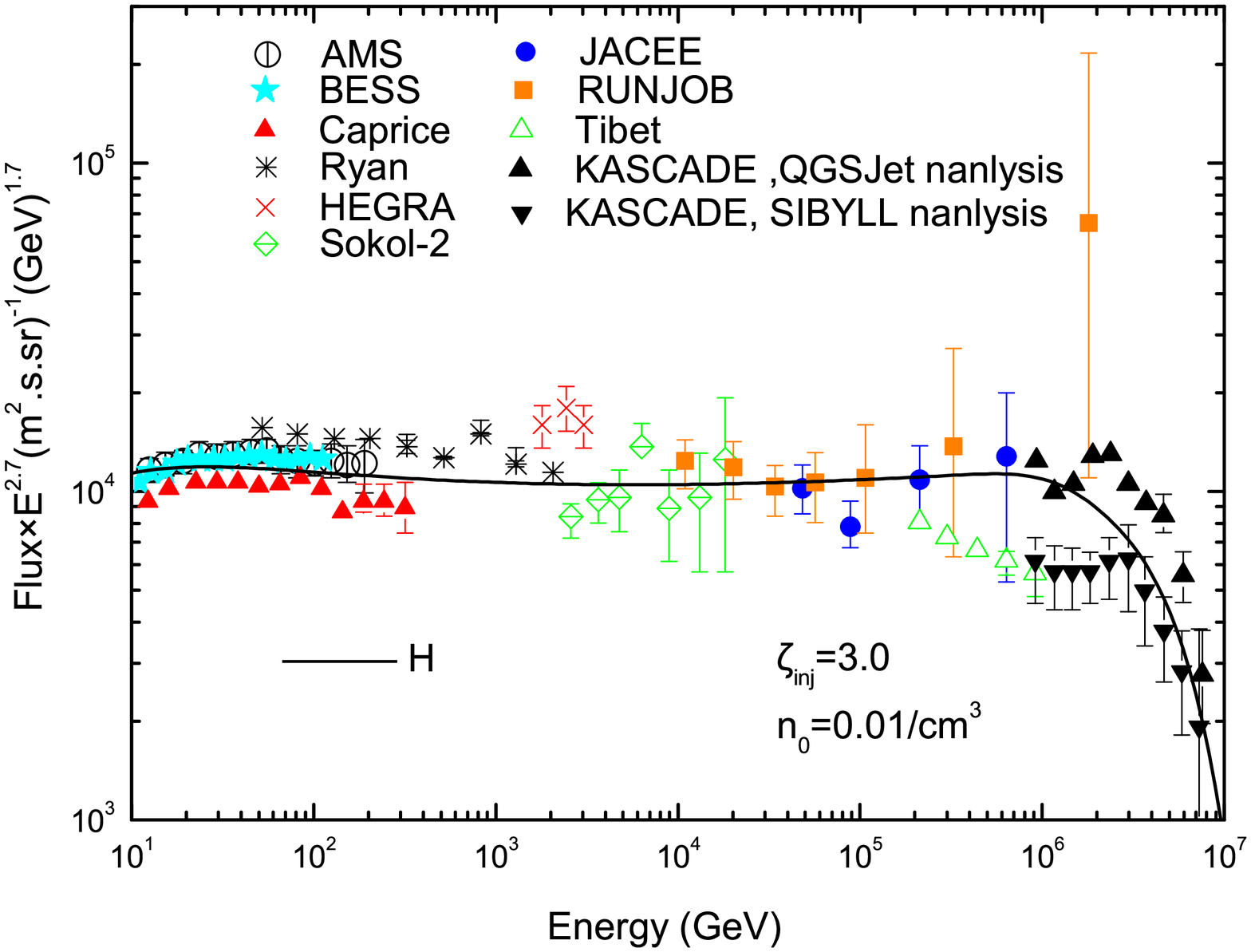}
\figcaption{\label{fig:4}  Comparison of model results with observed H flux. Solid line is H flux predicted in the model. Model parameters are  $\xi_{\rm H} = 3.5$, $\delta =0.5$, $\chi=0.5$, $T_0 = 10^6$ K, $n_0 = 0.01$ cm$^{-3}$, and $B_0=5~\mu$G. }
\end{center}

As mentioned above, the acceleration efficiency {plays} an important role for the cut-off energy of the spectrum in the model. KASCADE experiments \cite{lab21} show that the observed H spectrum has a knee feature at a few PeV, which is not consistent with that observed by ARGO-YBJ experiment \cite{lab2,lab14,lab15}. To reproduce the observed H flux by KASCADE experiments with our model, the acceleration efficiency ( i.e. $\xi_{\rm H}$) is properly adjusted but other {parameters} are not changed for the proton injection spectrum, and the comparison of model result with the observed H flux is shown in Fig.~\ref{fig:4}. {With ${\xi_{\rm H}=3.5}$ and ${\delta=0.5}$, the observed H flux can be reproduced well in this model.}

\section{Results and Discussion}

In this paper, the observed spectra of the light component (P+He) by ARGO-YBJ and the proton spectrum by KASCADE experiments are reproduced in the frame of non-linear diffusive shock acceleration model of the SNR. In the model, the escape spectrum of $i$th particles injected into the ISM consists of three components (see Eqs.  (\ref{eq:phiesc1}),  (\ref{eq:phiesc2}) and (\ref{eq:phiesc3})). {In our calculation, for the case with ARGO-YBJ, parameter ${\xi_{H}=3.8}$ is used, which corresponds to ${\eta_{H}\approx 6.5\times 10^{-5}}$ and ${\eta_{\rm He}\approx 0.31 K_{\rm He~H}\times\eta_{\rm H}\sim 2.01\times 10^{-5}}$, and ${\eta_{H}\approx 4.32\times 10^{-4}}$ for the case with KASCADE.} These values {indicate} that the acceleration at the SNR shock is {very dfficient}. In fact, it is generally believed that the acceleration is inefficient when $\eta_{\rm H}<\sim 10^{-5}$, which implies the acceleration efficiency is less than a few $\%$ \cite{lab8}.

Finally, it should be pointed out that a typical {evolution scenario} of SNRs in our Galaxy and a simple leaky box approximation of CR propagation are used in our calculations. In fact, the case is very complicated \cite{lab7} and the detailed {processes} of CR spectrum reaching the knee through NLDSA remain uncertain. Meanwhile, the variations of parameters have great influence on the distribution of the knee. The contributions of different classes of SNRs to the CR spectrum should be taken into account \cite{lab24} and more relativistic CR propagation model \cite{lab25} should be used. It can be predicted that the data with smaller uncertainty will constrain the parameters (e.g the amplified magnetic field) and demonstrate the effectiveness of the model.

\acknowledgments{The authors would like to thank the anonymous referee for insightful comments. We also thank Prof. Xiaohui Sun for helpful discussions and  suggestions.}

\end{multicols}

\vspace{10mm}

\vspace{-1mm}
\centerline{\rule{80mm}{0.1pt}}
\vspace{2mm}

\begin{multicols}{2}

\end{multicols}

\clearpage
\end{CJK*}
\end{document}